\documentclass[sigconf]{acmart}

\usepackage[utf8]{inputenc}
\usepackage{tabularray}
\usepackage{booktabs}
\usepackage{multirow}
\usepackage{makecell}
\usepackage{float}
\usepackage{color}
\setcopyright{none}
\renewcommand\footnotetextcopyrightpermission[1]{}
\AtBeginDocument{%
  }

\setcopyright{acmlicensed}
\copyrightyear{2018}
\acmYear{2018}
\acmDOI{XXXXXXX.XXXXXXX}

\acmConference[Conference acronym 'XX]{Make sure to enter the correct
  conference title from your rights confirmation emai}{June 03--05,
  2018}{Woodstock, NY}
\acmISBN{978-1-4503-XXXX-X/18/06}

\begin{document}

\title{Bridging Domain Gaps between Pretrained Multimodal Models and Recommendations}

\author{Wenyu Zhang}
\authornote{Co-first authors with equal contribution to refining the theory and experimental design.}
\orcid{0009-0001-1457-1707}
\affiliation{%
  \institution{University of Science and Technology of China}
  \city{Hefei}
  \country{China}}
\email{wenyuz@mail.ustc.edu.cn}

\author{Jie Luo}
\authornotemark[1]

\affiliation{%
  \institution{University of Science and Technology of China}
  \city{Hefei}
  \country{China}}
\email{luojie2000@mail.ustc.edu.cn}

\author{Xinming Zhang}
\authornote{Corresponding authors.}

\orcid{0000-0002-8136-6834}
\affiliation{%
  \institution{University of Science and Technology of China}
  \city{Hefei}
  \country{China}}
\email{xinming@ustc.edu.cn}

\author{Yuan Fang}
\authornotemark[2]
\affiliation{%
  \institution{Singapore Management University}
  \country{Singapore}}
\email{yfang@smu.edu.sg}








\renewcommand{\shortauthors}{Zhang et al.}


\begin{abstract}
With the explosive growth of multimodal content online, pre-trained visual-language models have shown great potential for multimodal recommendation. However, while these models achieve decent performance when applied in a frozen manner, surprisingly, due to significant domain gaps (e.g., feature distribution discrepancy and task objective misalignment) between pre-training and personalized recommendation, adopting a joint training approach instead leads to performance worse than baseline. Existing approaches either rely on simple feature extraction or require computationally expensive full model fine-tuning, struggling to balance effectiveness and efficiency. To tackle these challenges, we propose \textbf{P}arameter-efficient \textbf{T}uning for \textbf{M}ultimodal \textbf{Rec}ommendation (\textbf{PTMRec}), a novel framework that bridges the domain gap between pre-trained models and recommendation systems through a knowledge-guided dual-stage parameter-efficient training strategy. In the first stage, we leverage features extracted from pre-trained vision-language models to facilitate multimodal recommendation training, which enables the recommendation model to adapt to the representation space of pre-trained features while allowing ID embeddings to capture personalized user-item matching patterns. This knowledge is then leveraged in the second stage to guide parameter-efficient fine-tuning of the pre-trained model, thereby achieving effective domain adaptation while maintaining computational efficiency. This framework not only eliminates the need for costly additional pre-training but also flexibly accommodates various parameter-efficient tuning methods. Experiments on three public datasets demonstrate that PTMRec significantly improves recommendation performance (average 10.6\% gain in Recall@10) by training only x\% of parameters compared to the original CLIP model. Our work provides a practical and general framework for addressing domain adaptation challenges in multimodal recommendation.
\end{abstract}



\maketitle

\section{Introduction}
Recommender systems play a crucial role in personalized services \cite{50DBLP:journals/csur/ZhangYST19}, with traditional ID-based methods\cite{1_koren2009matrix,2_BPR,3_lightgcn} achieving significant success. However, as online platforms evolve \cite{51DBLP:conf/kdd/YingHCEHL18}, the prevalence of multimodal data challenges conventional approaches \cite{10_DBLP:journals/corr/abs-2302-04473}, driving the development of multimodal recommendation systems \cite{5_DBLP:conf/icmlc/ChenWHT17,6_DBLP:conf/aaai/FuPWXL19,52DBLP:conf/ijcai/DuLLZ22,56DBLP:journals/corr/abs-2411-10332,57DBLP:journals/corr/abs-2405-15304}. Recent advances in multimodal recommendation systems have spawned various approaches utilizing pre-trained models. From the perspective of leveraging pre-trained models, we categorize existing work into three paradigms: (1) Pre-extracted, (2) Pre-train/Fine-tuning, and (3) Pre-extracted with Parameter-Efficient Fine-tuning (PEFT).

\begin{figure}[t]
  \centering
   \includegraphics[width=\linewidth]{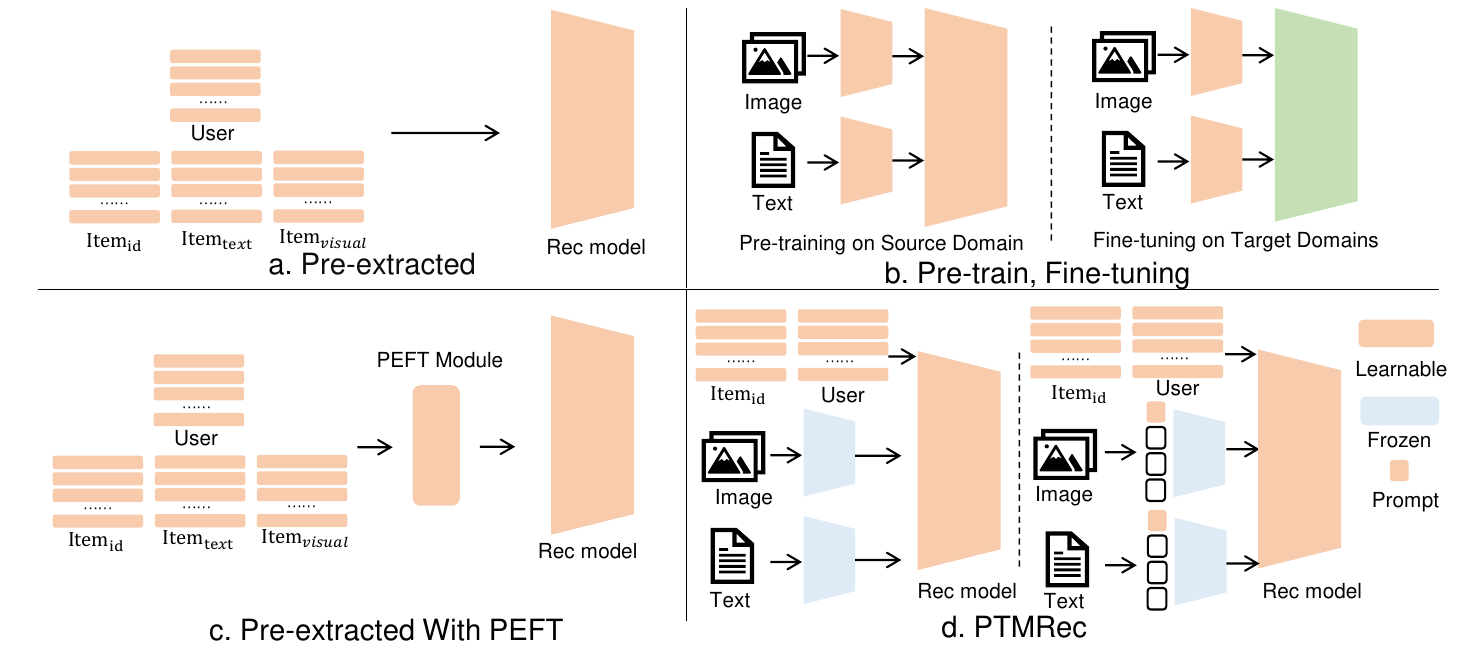}
   \caption{Modality Feature Extraction Paradigms in Multimodal Recommendation.}
   \label{fig:1}
\end{figure}

First, the pre-extracted paradigm uses pre-trained encoders to extract features offline, which, as shown in Fig.~\ref{fig:1}a, are used to initialize the feature embedding layer of recommendation models. VBPR \cite{7_DBLP:conf/aaai/HeM16} pioneered the introduction of pre-trained CNN features, MMGCN \cite{wei2019mmgcn} enhanced feature interaction through modality-specific graph structures, and Freedom \cite{15_DBLP:conf/mm/ZhouS23} improved performance via contrastive learning and cross-modal attention. However this paradigm mainly relies on early lightweight models (such as VGG-16 \cite{58DBLP:journals/corr/SimonyanZ14a} and Sentence-BERT \cite{18_DBLP:conf/emnlp/ReimersG19}), and the static nature of features limits the potential of recommendation models.

Second, as shown in Fig.~\ref{fig:1}b, the pre-training/fine-tuning paradigm employs pre-trained models such as BERT\cite{19_DBLP:conf/naacl/DevlinCLT19}, ViT~\cite{20_DBLP:conf/iclr/DosovitskiyB0WZ21}, and CLIP~\cite{21_radford2021learning} as modal encoders and fine-tunes them during the recommendation training process \cite{62yang2024exploring}. One line of work directly fine-tunes pre-trained models as modal encoders (e.g., MoRec~\cite{4_DBLP:conf/sigir/YuanYSLFYPN23}). Another line of work addresses domain transfer through source domain pre-training and target domain fine-tuning (e.g., TransRec~\cite{54DBLP:conf/apweb/WangYCJYKWHL24} and PMMRec~\cite{55DBLP:conf/icde/LiDNZGY024}). Notably, AdapterRec~\cite{25_DBLP:conf/wsdm/FuY0YCCZWP24} incorporates parameter-efficient adapters~\cite{27_DBLP:conf/icml/HoulsbyGJMLGAG19} during target domain fine-tuning to improve efficiency. Direct fine-tuning methods inevitably require substantial computational resources \cite{63DBLP:journals/pami/YangZC22}, while domain transfer approaches demand significant resources particularly in the source domain pre-training phase.

Third, the pre-extracted with PEFT paradigm adapts to recommendation scenarios through parameter-efficient methods such as prompts \cite{26_DBLP:conf/acl/LiL20}, adapters \cite{27_DBLP:conf/icml/HoulsbyGJMLGAG19}, and LoRA \cite{28_DBLP:conf/iclr/HuSWALWWC22}. As shown in Fig~\ref{fig:1}c, this paradigm primarily enhances pre-extracted features by introducing lightweight trainable modules. For instance, PromptMM \cite{22_DBLP:conf/www/WeiTXJH24} employs trainable prompt tokens and knowledge distillation to bridge modality content and collaborative signals, while MISSRec \cite{24_DBLP:conf/mm/WangZWWLLYZZX23} adopts multimodal adapters to enhance sequence representation. Furthermore, if PEFT methods can be effectively utilized to transfer powerful pre-trained models to the recommendation domain, it would significantly enhance the performance of recommendation models. However, our experiments show that directly applying PEFT methods for joint training of pre-trained encoders and recommender systems leads to performance degradation (detailed results in Section 3.3 Ablation study), indicating that the substantial domain gap between pre-trained models and recommender systems cannot be effectively bridged through traditional PEFT methods.

In summary, directly applying pre-trained models in existing recommender systems still faces key challenges: while \textbf{pre-extraction} and \textbf{PEFT-enhanced pre-extraction} methods are computationally efficient, they struggle to fully unleash the representational potential of pre-trained models; the \textbf{pre-train/fine-tuning} paradigm shows superior performance but comes with enormous computational costs. The core dilemma lies in: \textbf{how to construct a parameter-efficient paradigm that enables joint training of pre-trained models and recommender systems while maintaining efficiency?}

To address these limitations, we propose a Parameter-efficient Tuning for Multimodal Recommendation framework. As illustrated in Fig. \ref{fig:1}d, PTMRec introduces an innovative knowledge-guided dual-stage parameter-efficient training strategy that effectively bridges the domain gap between pre-trained models and recommendation systems while maintaining computational efficiency. In the first stage (Personalized Preference Learning), we train a lightweight recommendation model with learnable parameters using pre-trained features to capture task-specific knowledge about user preferences and item characteristics. In the second stage (Knowledge-guided Prompt Learning), we guide the tuning of pre-trained models through knowledge transfer optimization using personalized preference knowledge. This two-stage design not only eliminates the need for expensive additional pre-training but also provides a flexible framework that can accommodate various parameter-efficient tuning methods while maintaining the coupling between feature extraction and recommendation objectives.

\section{Method}

\subsection{Preliminaries}
Let $\mathcal{U} = {u_1, u_2, \ldots }$ denote the user set and $\mathcal{I} = {i_1, i_2, \ldots }$ denote the item set. For each user u, their historical interactions form an item set $\mathcal{I}^u$ representing positive feedback. Each item i contains visual and textual modalities ($\mathbf{X}_i^{\text{v}}$ and $\mathbf{X}_i^{\text{t}}$) besides ID information. For representations, we use embeddings: $\mathbf{e}_u^{\text{id}} \in \mathbb{R}^{d}$ and $\mathbf{e}_i^{\text{id}} \in \mathbb{R}^{d}$ for user and item IDs respectively, where d is the embedding dimension. Item multimodal information is represented by $\mathbf{e}_i^{\text{v}}$ and $\mathbf{e}_i^{\text{t}}$ from visual and textual features.

For optimization, recommendation systems typically employ BPR loss:

\begin{equation}
\mathcal{L}_{BPR} = -\sum_{(u,i,j)\in\mathcal{D}} \ln \sigma(f_u(i) - f_u(j))
\end{equation}

where $\sigma(\cdot)$ is the sigmoid function, $(u,i,j)$ represents a training triplet with user u, interacted item i, and non-interacted item j.

Vision-language pre-training typically adopts InfoNCE loss for image-text alignment:
\begin{equation}
\mathcal{L}_{NCE} = -\frac{1}{N}\sum_{i=1}^{N} \log\frac{\exp(sim(\mathbf{v}_i, \mathbf{t}_i)/\tau)}{\sum_{j=1}^{N}\exp(sim(\mathbf{v}_i, \mathbf{t}_j)/\tau)}
\end{equation}

where $\mathbf{v}_i$ and $\mathbf{t}_i$ are visual and textual features, $sim(\cdot,\cdot)$ is cosine similarity, and $\tau$ is temperature. These objectives reflect different goals: BPR focuses on learning personalized preferences through user-item interactions, while InfoNCE aims for general vision-language alignment.

\begin{figure*}[h!]
  \centering
  \includegraphics[scale=0.5]{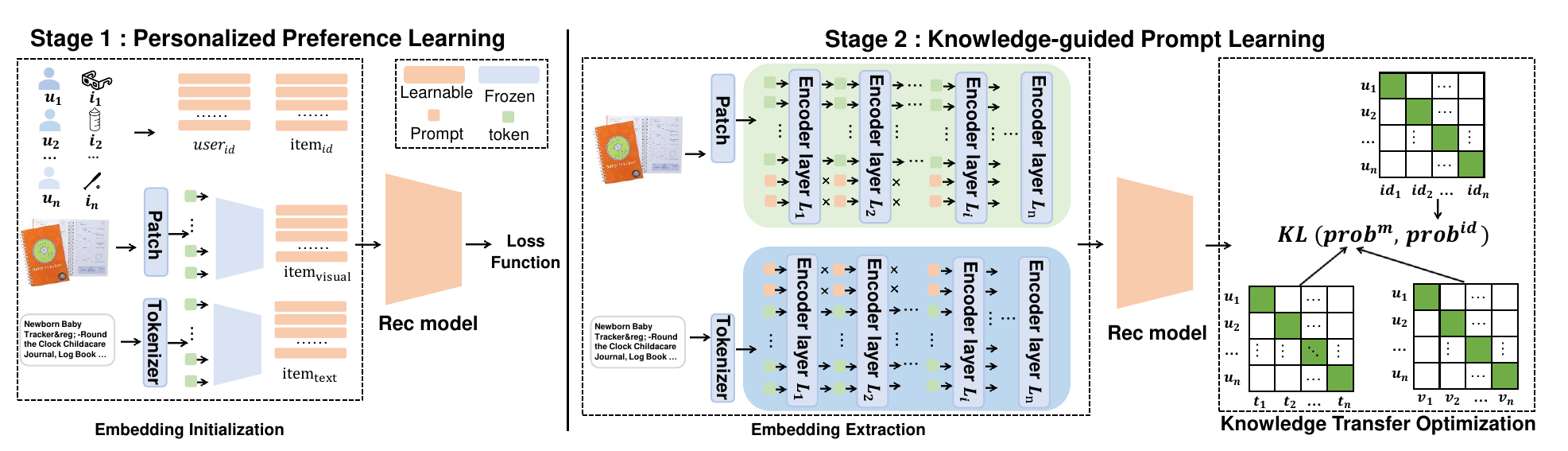} 

  \caption{The architectures of PTMRec.}
  \label{fig:3}
\end{figure*}

\subsection{Personalized Preference Learning}

As shown in Figure \ref{fig:3}, in the first stage of our framework, we aim to learn basic recommendation patterns while preserving the general semantic knowledge from pre-trained models. Specifically, we utilize a frozen CLIP model for multimodal feature extraction to initialize and train the recommendation model.

For each item i, we first extract its visual and textual features using the pre-trained CLIP model, where $\mathbf{X}_i^{\text{v}}$ and $\mathbf{X}_i^{\text{t}}$ denote the raw image and text description of item i respectively. The image encoding divides the image $\mathbf{X}_i^{\text{v}}$ into $M$ fixed-size blocks, which are projected as image block embeddings 
$\mathbf{E}^{\text{v}} =[\text{e}_0^{\text{v}},\text{e}_1^{\text{v}}, \cdots,\text{e}_N^{\text{v}}]$. A learnable class (CLS) token $\mathbf{c}$ is initialized and input into the image encoder together with the image block embeddings. The image representation $\mathbf{e}_i^{\text{v}}$ is obtained by mapping the class token $\mathbf{c}'$ from the last Transformer layer through $\texttt{VisualProj}$:
\begin{equation} [{\mathbf{c'}},\mathbf{E'}^{\text{v}}]=\text{VisualEnc}([\mathbf{c},\text{E}^{\text{v}}]). \end{equation}
\begin{equation} \mathbf{e}_i^{\text{v}}=\text{VisualProj}({\mathbf{c'}}). \end{equation}

The CLIP text encoder uses a tokenizer to project the text into token $\text{E}^{\text{t}} =[\text{e}_0^{\text{t}},\text{e}_1^{\text{t}}, \cdots,\text{e}_N^{\text{t}}]$. This is input into the text encoder \text{TextEnc}. The final text representation $\mathbf{e}_i^{\text{t}}$ projects the last token output from the last Transformer layer through $\texttt{TextProj}$ into the general multimodal embedding space.
\begin{equation} \mathbf{E'}^{\text{t}}=\text{TextEnc}(\mathbf{E}^{\text{t}}). \end{equation}
\begin{equation} \mathbf{e}_i^{\text{t}}=\text{TextProj}({\mathbf{e'}_{N}^{\text{t}}}). 
\end{equation}

These encoders are kept frozen during training. Different recommendation models can be flexibly integrated into our framework and trained with their own specific objective functions.

\subsection{Knowledge-guided Prompt Learning}

Building upon the basic recommendation patterns learned in the first stage, we design a knowledge-guided parameter-efficient tuning strategy to enhance model performance. Our framework is flexible and can incorporate various parameter-efficient methods (e.g., LoRA, Adapter, or other methods). Through extensive experimental (Table \ref{tab:PEFT}) analysis balancing computational efficiency and model performance, we find that prompt learning is the most suitable choice for recommendation scenarios. Therefore, we focus on achieving efficient domain adaptation by introducing learnable prompts into CLIP encoders.

As shown in Fig.~\ref{fig:3}, we follow the design of MaPLe \cite{29_DBLP:conf/cvpr/KhattakR0KK23} and insert visual and textual prompts, denoted as $p^m=\langle{p^m_{0},p^m_{1},\cdots,p^m_n}\rangle $, into the first i layers of the modality encoder, where $ m \in (t,v)$. Different from MaPLe, we remove the prompt coupling function for modality alignment since recommendation systems focus more on personalized perception.
\begin{equation}
[{\mathbf{c'}}, \mathbf{E}_l^{\prime \text{v}}, \_] = \text{VisualEnc}_l([\mathbf{c_l},\text{E}_l^{\text{v}}, p^v_l]).   
 ~  l=1, \cdots, i \\
\end{equation}
\begin{equation} 
[{\mathbf{c}^{\prime}},\mathbf{E}_{l+1}^{\prime \text{v}}, p^v_{l+1}] = \text{VisualEnc}_l
([\mathbf{c_{l+1}},\text{E}_{l}^{\text{v}}, p^v_l]).   ~  l=i+1, \cdots, L 
\end{equation}

For the text encoder, it select the last token as the text representation. We insert $p^t$ before the text tokens.
\begin{equation}
[\_, \mathbf{E'}^{\text{t}}] = \text{TextEnc}_l([p^t_l,\text{E}_l^{\text{t}}]).  ~  l=1, \cdots, i 
\end{equation}

\begin{equation}
[p^\top_{l+1},\mathbf{E'}_{l}^{\text{t}}] = \text{TextEnc}_l([p^t_l, \text{E}_l^{\text{t}}]).  ~  l=i+1, \cdots, L 
\end{equation}

In the first i layers of the modality encoder, the modality prompt is only referenced within the current layer, and a new prompt is introduced for each layer. All prompts are initialized using the standard random initialization method. After the i-th transformer layer, the subsequent layers process the output of the previous layer. 

\subsection{Knowledge Transfer Optimization}

To bridge the domain gap between pre-trained models and recommendation models, we propose an optimization strategy based on in-batch knowledge transfer. Specifically, in each training batch, we construct positive user-item pairs from interaction records while treating other items within the batch as negative samples. The user-item ID interaction distributions learned in the first stage contain core recommendation patterns. Through in-batch knowledge transfer, we effectively transfer these patterns to modal feature learning without additional computational overhead. The target distribution is defined as:
\begin{equation}
\text{probs}^{id} = \text{stopgrad}(\text{softmax}(\mathbf{e^u}^\top \cdot \mathbf{e}_{id})),
\end{equation}
where $\mathbf{e}^\top_u \cdot \mathbf{e}_{id}$ computes similarity scores between user and item features, which are normalized into probability distributions through softmax. The stopgrad operation prevents gradients from flowing through ID features, thus maintaining a stable target distribution and avoiding representation collapse \cite{60DBLP:conf/cvpr/ChenH21}.

For modal features, we compute their interaction probability distributions as follows:
\begin{equation}
\text{prob}^{m} = \text{log\_softmax}(\mathbf{e^u}^\top \cdot \text{Linear}(\mathbf{e}^m)), \quad \text{where} \ m \in {(t, v)},
\end{equation}
where we obtain dimension-aligned feature representations by applying linear transformation to modal features $\mathbf{e}^m$. Then, we minimize KL divergence to guide modal features in learning user-item interaction patterns. The specific computation is as follows:
\begin{equation}
\mathcal{L}_{KT} = \text{KL}(\text{prob}^t, \text{prob}^{id}) + \text{KL}(\text{prob}^v, \text{prob}^{id}).
\end{equation}

\begin{table*}[t]
	\caption{Overall performance comparison by different recommendation methods in terms of Recall and NDCG. Results are averaged over 5 runs with different random seeds. $^{*}$  indicates statistical significance (p < 0.05).}
	\footnotesize
	\label{Table:perform}
	\begin{center}
{
	\begin{tabular}{ccccccccccccc}
		\toprule
		\multirow{2}{*}{\textbf{Dataset}} & \multirow{2}{*}{\textbf{Metric}} & \multicolumn{3}{c}{\textbf{General models}} & \multicolumn{8}{c}{\textbf{Multi-modal models}} \cr
			\cmidrule(lr){3-5} \cmidrule(lr){6-7} \cmidrule(lr){8-9} \cmidrule(lr){10-11} \cmidrule(lr){12-13}  & & \textbf{BPR} & \textbf{LightGCN} & \textbf{BUIR} & \textbf{VBPR} & \textbf{$\text{VBPR}_{\text{PTM}}$}   & \textbf{Freedom} & \textbf{$\text{Freedom}_{\text{PTM}}$}  & \textbf{MGCN} & \textbf{$\text{MGCN}_{\text{PTM}}$} & \textbf{$\text{SMORE}$}  & \textbf{$\text{SMORE}_{\text{PTM}}$} \\
            \hline
			\multirow{4}{*}{Baby} 
                 & Recall@10 & 0.0357         & 0.0479   & 0.0506 & 0.0412             & \textbf{0.0532$^{*}$}    & 0.0646  & \textbf{0.0686$^{*}$}   & 0.0638 & \textbf{0.0649} &0.0678   &\textbf{0.0688$^{*}$}  \\
                 & Recall@20 & 0.0575         & 0.0754   & 0.0788 & 0.0672             & \textbf{0.0817$^{*}$ }  & 0.0982  & \textbf{0.1042$^{*}$}     & 0.0982 &\textbf{ 0.1014$^{*}$}  &0.1030  &\textbf{0.1060$^{*}$} \\
                 & NDCG@10   & 0.0192         & 0.0257   & 0.0269 & 0.0226             & \textbf{0.0283$^{*}$ }  & 0.0349  & \textbf{0.0369$^{*}$}   & 0.0344 & \textbf{0.0352$^{*}$}  &0.0371    &\textbf{0.0375$^{*}$}  \\
                 & NDCG@20   & 0.0249         & 0.0328   & 0.0342 & 0.0293             &\textbf{ 0.0356$^{*}$}    & 0.0436  & \textbf{0.0461$^{*}$}    & 0.0433 & \textbf{0.0446$^{*}$}  &0.0461  &\textbf{0.0470$^{*}$} \\
			\hline
			\multirow{4}{*}{Sports} 
                 & Recall@10 & 0.0432         & 0.0569   & 0.0467 & 0.0534             & \textbf{0.0582$^{*}$}    & 0.0715  & \textbf{0.0737$^{*}$}    & 0.0745 & \textbf{0.0762$^{*}$}  &0.0751 &\textbf{0.0764$^{*}$}  \\
                 & Recall@20 & 0.0653         & 0.0864   & 0.0733 & 0.0813             & \textbf{0.0896$^{*}$ }   & 0.1081  & \textbf{0.1107$^{*}$}     & 0.1108 & \textbf{0.1125$^{*}$}  &0.1122 &\textbf{0.1142$^{*}$} \\
                 & NDCG@10   & 0.0241         & 0.0311   & 0.0260  & 0.0286             & \textbf{0.0308$^{*}$}   & 0.0388  & \textbf{0.0396$^{*}$}     & 0.0409 & \textbf{0.0419$^{*}$}  &0.0412 &\textbf{0.0419$^{*}$} \\
                 & NDCG@20   & 0.0298         & 0.0387   & 0.0329 & 0.0358             & \textbf{0.0389$^{*}$}   & 0.0482  & \textbf{0.0491$^{*}$}   & 0.0502 & \textbf{0.0513$^{*}$}  &0.0508 &\textbf{0.0516$^{*}$} \\
			\hline
			\multirow{4}{*}{Clothing} 
                 & Recall@10 & 0.0235         & 0.0363   & 0.0332 & 0.0302             &\textbf{ 0.0457$^{*}$}   & 0.0629  & \textbf{0.0655$^{*}$ }   & 0.0665 & \textbf{0.0677$^{*}$}  &0.0656 &\textbf{0.0674$^{*}$} \\
                 & Recall@20 & 0.0367         & 0.0540    & 0.0514 & 0.0444             & \textbf{0.0696$^{*}$ }  & 0.0929  & \textbf{0.0969$^{*}$}    & 0.0965 & \textbf{0.1000$^{*}$}  &0.0971 &\textbf{0.1003$^{*}$} \\
                 & NDCG@10   & 0.0127         & 0.0204   & 0.0185 & 0.0169            & \textbf{0.0242$^{*}$}   & 0.0342  & \textbf{0.0355$^{*}$}     & 0.0364 & \textbf{0.0367$^{*}$}  &0.0358 &\textbf{0.0367$^{*}$} \\
                 & NDCG@20   & 0.0161         & 0.0250    & 0.0232 & 0.0205             & \textbf{0.0302$^{*}$}    & 0.0418  & \textbf{0.0435$^{*}$ }   & 0.0440 & \textbf{0.0449$^{*}$}  &0.0438  &\textbf{0.0451$^{*}$} \\
		\bottomrule			
		
		\end{tabular}}
	\end{center}
\end{table*}

\section{Experiment}

\subsection{Experiment setting}

\subsubsection{Baseline}
To validate our framework's effectiveness, we conduct comparative experiments with two categories of representative models: traditional recommendation models including BPR (UAI'09) \cite{2_BPR}, LightGCN (SIGIR'20) \cite{3_lightgcn}, BUIR (SIGIR'21) \cite{DBLP:conf/sigir/LeeKJPY21}, and multimodal recommendation models including VBPR (AAAI'16) \cite{7_DBLP:conf/aaai/HeM16}, Freedom (MM'23) \cite{15_DBLP:conf/mm/ZhouS23}, MGCN (MM'23) \cite{16_DBLP:conf/mm/Yu0LB23}, SMORE (WSDM'25) \cite{61ong2024spectrum}, comparing their performance both in original form and after integration with our framework.

\subsubsection{Implementation details.}
We selected three categories from the Amazon review dataset: Baby, Sports, and Clothing. All data is preprocessed through the MMRec framework \cite{10_DBLP:journals/corr/abs-2302-04473} and image is download from link. Consistent with existing models, We sets the embedding dimensions for users and items to 64 and employs the Adam optimizer with a learning rate of 0.001. To ensure model convergence during training, the number of epochs is set to 1000, and an early stopping strategy. The batch size for the first stage is set to 2048 (128 for the second stage with gradient accumulation steps of 12). The model evaluation was conducted on an NVIDIA Tesla V100 32 GB GPU.

\subsection{Performance Comparison}
As shown in Tab. \ref{Table:perform}, we compared conventional recommendation methods with popular multimodal recommendation methods across three datasets. Consistent with existing multimodal recommendation studies, we adopted Recall@K and NDCG@K (K=10,20) as evaluation metrics. The \textbf{PTM} indicates that our framework was utilized in the model. The results reveal that significant improvements were achieved when the multimodal model was incorporated into our framework. The simpler the model, the greater the performance enhancement. Furthermore, the enhanced performance indicates that our framework not only improves the overall performance of the models but also narrows the performance gap between them. The experimental results validate our claim that our method can effectively facilitate the migration of basic models to recommendation domains with substantial domain gaps, solely with a limited number of prompts.

\subsection{ Model Analysis}

\begin{table}[]
\caption{Ablation study of PEFT Methods\textsuperscript{*}.}
\label{tab:PEFT}
\footnotesize
\begin{tabular}{llccccccc}
\toprule
Datasets & Method     & Recall@20  & Time/E\textsuperscript{1}     & Param & MU\textsuperscript{2} \\
\midrule
\multirow{3}{*}{Baby} 
         & adapter    & 0.1046       & 20.5m    & 169M  & 23g   \\
         & lora & 0.103     & 23.5m    & 160M  & 28g \\
         & prompt   &0.1042     & 8m    & 9M  & 12g \\

\midrule
\multirow{3}{*}{Sports}
         & adapter    & 0.1107       &38m    & 182M  & 23g   \\
         & lora & 0.1095     & 44m    & 173M  & 28g \\
         & prompt   &0.1107      & 12.5m    & 22M  & 12g \\

\midrule
\multirow{3}{*}{Clothing}
         & adapter    & 0.0953       & 34m    & 188M  & 23g   \\
         & lora & 0.0955   &39m    & 178M  & 28g \\
         & prompt   &0.0969     &11m    & 28M  & 12g \\

\bottomrule
\end{tabular}
\\
\textsuperscript{*}Time/E and MU denote training time per epoch  and memory usage, respectively.
\end{table}

\subsubsection{Analysis of PEFT Methods }
To select an appropriate Parameter-Efficient Fine-Tuning (PEFT) method, we conducted comparative experiments on model performance, computational efficiency, and memory consumption across three datasets. As shown in Table \ref{tab:PEFT}, Prompt Tuning demonstrates significant advantages in memory usage and computational efficiency while maintaining outstanding performance, making it our chosen implementation approach. The experimental results also verify that our framework can be flexibly integrated with different PEFT methods. (Note: Sports and Clothing datasets have more parameters than the Baby dataset due to their larger item counts.)

\begin{figure}[h]
  \vspace{-2mm}

  \centering
   \includegraphics[width=\linewidth]{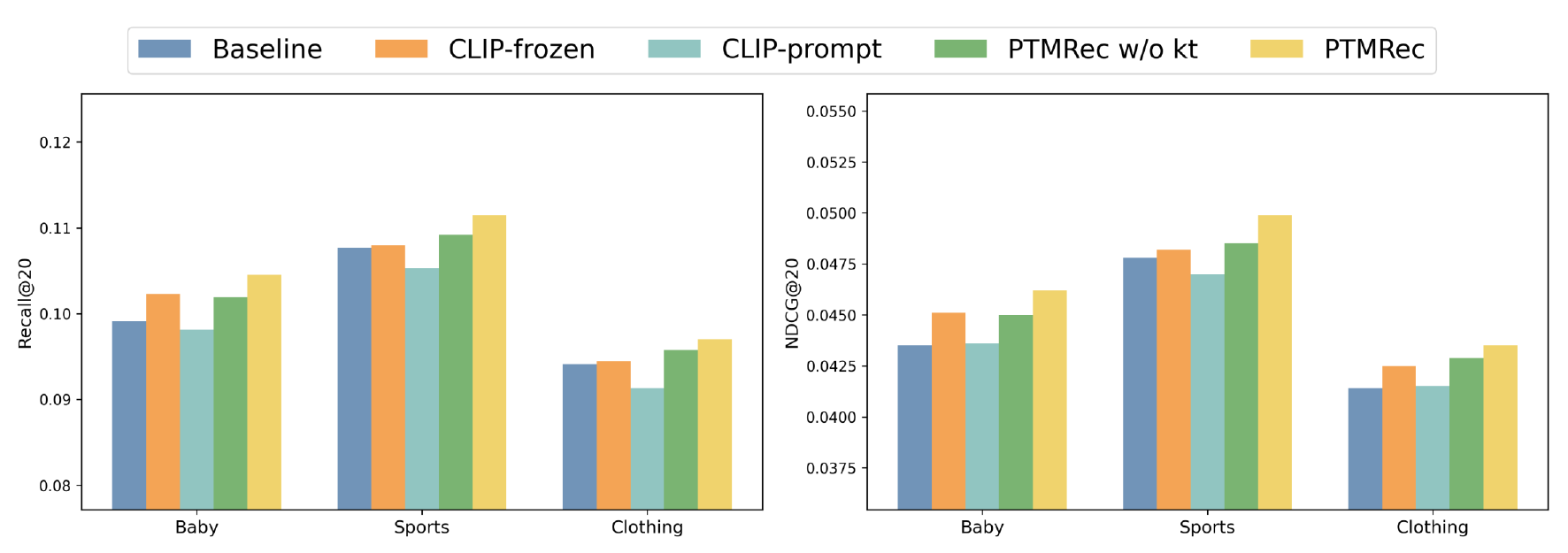}
   \caption{ Ablation study of PTMRec.}
   \label{fig:ablation}
\end{figure}

\subsubsection{Ablation Study }

To validate the contribution of each component, we conducted ablation studies based on the Freedom model across three datasets. Specifically, we tested variants including CLIP feature initialization (\textbf{CLIP-frozen}), joint training with CLIP pre-prompt (\textbf{CLIP-prompt}), two-stage training without knowledge transfer (\textbf{PTMRec w/o loss}), and the complete framework (\textbf{PTMRec}). As shown in Fig. \ref{fig:ablation}: (1) while CLIP-extracted modality features can enhance model performance, joint training with pre-prompt leads to performance degradation, indicating that simple prompts struggle to bridge the domain gap between recommendation and image-text matching; (2) although the two-stage training strategy partially alleviates the domain gap, optimizing prompts to obtain recommendation-suitable multimodal features remains challenging without domain knowledge guidance; (3) the complete framework, combining training decoupling and knowledge transfer loss, effectively mitigates domain differences, enabling the pre-trained model to transfer to domains with significant gaps.

\subsection{Visualization }
\begin{figure}[h]
  \vspace{-2mm}

  \centering
   \includegraphics[width=\linewidth]{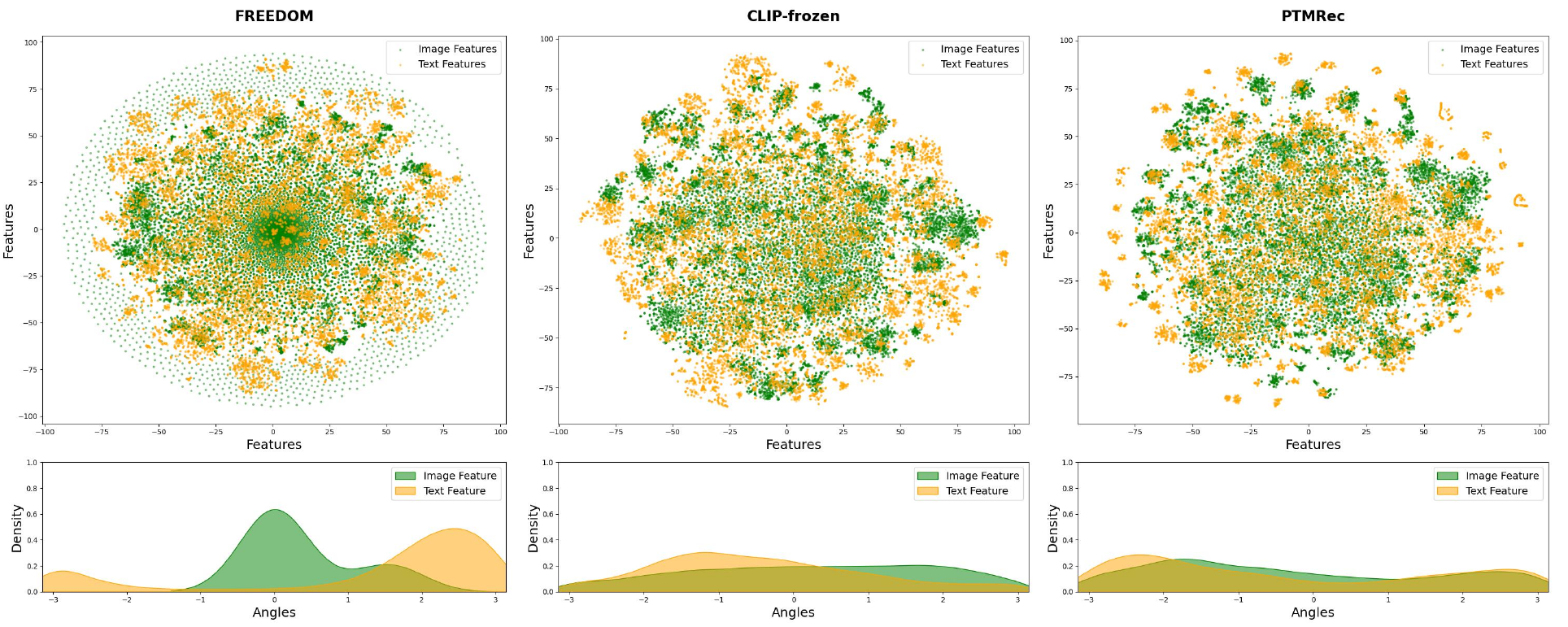}
   \caption{ Distribution of modal feature representations of Freedom in the Sports dataset under different settings. 
}
   \label{fig:visualize}
\end{figure}

We visualized feature distributions on the Amazon Sports dataset using t-SNE and Gaussian kernel density estimation (see Fig.~\ref{fig:visualize}). The results show that: the original Freedom model's features exhibit high clustering but significant dispersion, with dramatic fluctuations in density curves indicating poor feature alignment; after introducing the CLIP encoder, the feature distribution becomes smoother with enhanced local cohesion; furthermore, the complete PTMRec framework strengthens both local cohesion and multimodal alignment of features, leading to tighter clustering of similar items' features and thus improved recommendation performance.


\section{Conclusion}
This paper proposes PTMRec, a framework addressing the high training costs and domain gaps when integrating pretrained multimodal models into recommendation systems. Based on the CLIP modality encoder, we employ parameter-efficient tuning to reduce training expenses, and enhance user preference perception through a two-stage training strategy with knowledge transfer regularization.


\bibliographystyle{ACM-Reference-Format}
\bibliography{main}

\end{document}